\documentclass[12pt,a4paper]{article}
\usepackage{times}
\usepackage{a4wide}
\usepackage{amsfonts}
\usepackage{amssymb}
\usepackage{amsmath}
\usepackage{ifpdf}
\ifpdf
\usepackage[pdftex,implicit]{hyperref}
\hypersetup{%
  pdftitle    = {H-FGK formalism for black-hole solutions of N=2, d=4
    and d=5 supergravity},
  pdfkeywords = {supersymmetry, supergravity, special K\"{a}hler geometry, black holes, BPS solutions},
  pdfauthor   = {Patrick Meessen, Tom\'{a}s Ort\'{\i}n, Jan Perz, C. S. Shahbazi},
  plainpages  = true,
  colorlinks  = true,
  citecolor   = blue,
  urlcolor    = red,
  linkcolor   = black
}
\newcommand{\hepth}[1]{{\tt
\href{http://www.arXiv.org/abs/hep-th/#1}{hep-th/#1}}}

\newcommand{\arxiv}[1]{{\tt
\href{http://www.arXiv.org/abs/#1}{arXiv:#1}}}
\else
  \usepackage[dvips]{graphicx}
  \usepackage[unicode,implicit]{hyperref}
  \newcommand{\hepth}[1]{{\tt hep-th/#1}}

  \newcommand{\arxiv}[1]{{\tt arXiv:#1}}
\fi
\usepackage{tikz}
\newcommand{\FPAUO}[2]{
\tikz[scale=.13,
         Uniovi/.style={color=gray, fill=gray}
 ] {
 \fill[Uniovi] (0,0) circle (10);
 \fill[white] (0,7) circle (1.5);
 \draw[Uniovi] (-2,7.5) rectangle (2,5.5);
 \fill[white] (-0.3,6.6) rectangle (0.3,0);   
 \fill[white] ( -0.9,6.2) rectangle (.9 ,5.6);
 \fill[white] (-1.4, 5.2) rectangle (1.4, 4.6);
 \fill[white] (0,0) ellipse (3.5 and 4);
 \fill[Uniovi] (-2.5,0.3) rectangle (2.5,-0.3);
 \fill[Uniovi] (-2,2.3) rectangle (2,1.7);
 \fill[Uniovi] (-2,-2.3) rectangle (2,-1.7);
 \fill[white] (-4.5,5.5) rectangle (-2.7,4.9);
 \fill[white] (-3.9,6.1) rectangle (-3.3,4.3);
 \fill[white] (4.5,5.5) rectangle (2.7,4.9);
 \fill[white] (3.9,6.1) rectangle (3.3,4.3);
 \foreach \x in { 0,..., 3 }
   \foreach \y in { 0,...,\x}
    {
     \fill[white] (-6-\x*0.7+\y*1.4,3.5-\x *1.97) -- (-5.6-\x*0.7+\y*1.4,2.4-\x *1.97) -- (-6.4-\x*0.7+\y*1.4,2.4-\x *1.97) -- cycle;
     \fill[white] (6-\x*0.7+\y*1.4,3.5-\x *1.97) -- (5.6-\x*0.7+\y*1.4,2.4-\x *1.97) -- (6.4-\x*0.7+\y*1.4,2.4-\x *1.97) -- cycle;
   };
 \draw (0,-6) node[
                               text centered, 
                               color=white, 
                               font={\fontsize{8}{4}\sffamily\selectfont}
                             ] {FPAUO-#1/#2};
}} 
\makeatletter
\@addtoreset{equation}{section}
\makeatother

\begin{document}

~\vspace{-4cm}\begin{flushright}
\small
\FPAUO{11}{13}\\
IFT-UAM/CSIC-11-93\\
March 9\textsuperscript{th}, 2012\\
\normalsize
\end{flushright}

\vspace{2cm}

\begin{center}

{\Large {\bf H-FGK formalism for black-hole solutions}}\\[.5cm] 
{\Large {\bf of $N=2$, $d=4$ and $d=5$ supergravity}}

\vspace{2cm}

\renewcommand{\thefootnote}{\alph{footnote}}
{\sl\large Patrick Meessen$^{\dagger}$}
\footnote{E-mail: {\tt meessenpatrick@uniovi.es}},
{\sl\large Tom\'{a}s Ort\'{\i}n$^{\diamond}$}
\footnote{E-mail: {\tt Tomas.Ortin@csic.es}},
{\sl\large Jan Perz$^{\diamond}$}
\footnote{E-mail: {\tt Jan.Perz@uam.es}}
{\sl\large and C.~S.~Shahbazi$^{\diamond}$}
\footnote{E-mail: {\tt Carlos.Shabazi@uam.es}}
\setcounter{footnote}{0}
\renewcommand{\thefootnote}{\arabic{footnote}}

\vspace{.5cm}
${}^{\dagger}${\it  HEP Theory Group, Departamento de F\'{\i}sica, Universidad de Oviedo\\ 
        Avda.~Calvo Sotelo s/n, 33007 Oviedo, Spain}\\
\vspace{.4cm}
${}^{\diamond}${\it Instituto de F\'{\i}sica Te\'orica UAM/CSIC\\
C/ Nicol\'as Cabrera, 13--15,  C.U.~Cantoblanco, 28049 Madrid, Spain}\\
\vspace{2cm}
{\bf Abstract}
\end{center}
\begin{quotation}\small
  We rewrite the Ferrara-Gibbons-Kallosh (FGK) black-hole effective action of
  $N=2$, $d=4,5$ supergravities coupled to vector multiplets, replacing the
  metric warp factor and the physical scalars with real variables that
  transform in the same way as the charges under duality transformations, which
  simplifies the equations of motion. For a given model, the form of the
  solution in these variables is the same for all spherically symmetric black
  holes, regardless of supersymmetry or extremality.
\end{quotation}

\newpage
\pagestyle{plain}



\section*{Introduction}

During the past 20 years a huge effort has been made to classify, construct
and study black-hole solutions of 4- and 5-dimensional supergravity
theories. Most of this work has been devoted to the extremal black-hole
solutions and, in particular, to the supersymmetric ones. This was partly due
to their very special properties, such as their classical and quantum
stability, the attractor mechanism \cite{Ferrara:1995ih}, which makes their
entropies understandable and computable from a microscopic point of view
\cite{Strominger:1996sh}, and partly due to their functional simplicity, which
allows for the explicit and systematic construction of all of them (see {\it
  e.g.\/}~Ref.~\cite{Mohaupt:2000mj}).

The territory of non-extremal black holes, which includes as its boundary both
supersymmetric and non-supersymmetric extremal solutions, although potentially
more interesting, remains largely unexplored.\footnote{For some earlier works
  on non-extremal black-hole solutions see Refs.~\cite{Horowitz:1996fn}.}
Recently, in Refs.~\cite{Galli:2011fq,Meessen:2011bd}, we have attempted to
make the construction of non-extremal black-hole solutions of $N=2,d=4,5$
supergravity coupled to vector supermultiplets as systematic as that of the
extremal supersymmetric ones. The proposal is based on a deformation of
supersymmetric extremal solutions.

The fields of the supersymmetric black-hole solutions of $N=2,d=4$ and $d=5$
supergravity coupled to $n$ vector supermultiplets are given in terms of
functions $H^{M}$, $M=1, \dotsc, 2n+2$ and $H_{I}$, $I=1,\dotsc, n+1$,
respectively, that transform linearly under the generic duality groups
$Sp(2n+2,\mathbb{R})$ and $SO(n+1)$. The equations of motion are satisfied when
these functions are harmonic on the transverse $\mathbb{R}^{d-1}$ space (see
Refs.~\cite{Behrndt:1996jn,Meessen:2006tu} and \cite{Gauntlett:2002nw}). The
functional dependence of the physical fields on the variables $H$ is
essentially unique, since the linear action of the duality groups on the $H$'s
must produce a fixed non-linear transformation of the scalars and leave the
spacetime metric invariant. It is therefore natural to expect that all
black-hole solutions of the same model have the same functional form in terms
of the variables $H$, although in the non-extremal case or with gaugings these
variables will have to satisfy different equations and will no longer be
harmonic.\footnote{Here we will only deal with the ungauged cases and we will
  not include hypermultiplets, but experience shows that in the gauged cases
  \cite{Bellorin:2007yp,Huebscher:2007hj,Cacciatori:2008ek,Bellucci:2008cb,Dall'Agata:2010gj}
  and in the cases with hyperscalars \cite{Huebscher:2006mr,Bellorin:2006yr}
  the $H$'s always appear in the same way in the physical fields. In the cases
  with hyperscalars the $H$-functions are still harmonic, but on spaces more
  general than $\mathbb{R}^{d-1}$.}  A proposal made and checked for several
models in Refs.~\cite{Galli:2011fq,Meessen:2011bd} suggests that for
non-extremal black holes $H$ could be exponential or hyperbolic functions, but
it is mandatory to examine both the universality of solutions expressed in $H$
and of the hyperbolic ansatz.

The formalism introduced by Ferrara, Gibbons and Kallosh (FGK) in
Ref.~\cite{Ferrara:1997tw}, and generalized to arbitrary dimensions in
Ref.~\cite{Meessen:2011bd}, provides us with a convenient setting to
investigate these issues. Thanks to a suitable choice of space-time
coordinates for spherically symmetric black holes and to the replacement of
the vector fields by the charges, the bosonic sector of the supergravity
action is reduced to an effective action in ordinary mechanics, whose only
remaining dynamical variables are the scalar fields and the metric function
$U$; the role of the evolution parameter is played by a radial
coordinate. Black holes correspond to stationary points of this effective
action.

To answer the first question, in this Letter we adapt the FGK formalism by
rewriting the FGK effective action in terms of the variables $H$, which will
be the new degrees of freedom. The field redefinition is modelled on the
functional form of the extremal supersymmetric black-hole solutions, but
depends neither on supersymmetry nor extremality. The fact that such a change
of variables is possible means that, as functions of the $H$'s, all black-hole
solutions of a given model have the same form as the supersymmetric
solution. The action, expressed in the new variables, takes a particularly
simple $\sigma$-model form with the same Hessian metric\footnote{ A metrical
  manifold is said to be a Hessian manifold if there exists a coordinate
  system in which the components of the metric can be obtained as the Hessian
  of some function. Hessian metrics appear quite naturally in supergravity
  theories \cite{Mohaupt:2010fk,kn:MV} and also in global supersymmetry
  \cite{VandenBleeken:2011ib}. } occurring both in the kinetic term for the
scalars and in the effective potential generated by the gauge fields. For the
5-dimensional case, following a different, more general, formalism developed
in Ref.~\cite{Mohaupt:2009iq}, the same results have recently been obtained in
Ref.~\cite{Mohaupt:2010fk}. The analogous results for the 4-dimensional case
are presented here for the first time.\footnote{When the present work was being
  readied for publication, Ref.~\cite{kn:MV} appeared, with similar results
  for the 4-dimensional case in a more general framework.}

Having derived the equations of motion for the variables $H$, we can address
the second problem: Are all extremal black-hole solutions given by harmonic
functions, also in the non-supersymmetric case? Are the non-extremal black-hole
solutions always given by hyperbolic functions? This will be the subject of a
forthcoming publication \cite{kn:us}.


\section{H-FGK for \texorpdfstring{$N=2$}{N=2}, \texorpdfstring{$d=5$}{d=5} supergravity}
\label{sec-d5}

We find it more natural to present our proposal first in five dimensions. In
$N=2$, $d=5$ supergravity coupled to $n$ vector multiplets the physical fields
defining a black-hole solution with given electric charges $q_{I}$
($I=0,\dotsc, n$) are the metric function $U$ and the $n$ real scalars
$\phi^{x}$. Through a field redefinition we will replace them by $n+1$
variables denoted by $H_{I}$. We will also define a set of $n+1$ dual
variables $\tilde{H}^{I}$, which are useful for intermediate steps in our
calculation, but can also be used for finding other kinds of solutions
\cite{Martin:2012bi}.

The starting point\footnote{ Our conventions are those of
  Refs.~\cite{Bergshoeff:2004kh,Bellorin:2006yr}. } is the function
$\mathcal{V}(h^{\cdot})$, homogeneous of third degree in $n+1$ variables
$h^{I}$,
\begin{equation}
\mathcal{V}(h^{\cdot}) \equiv C_{IJK}h^{I}(\phi)h^{J}(\phi)h^{K}(\phi)\, ,  
\end{equation}
\noindent
which defines the scalar manifold as the hypersurface $\mathcal{V} = 1$. The
dual scalar functions $h_{I}$ are defined in terms of the $h^{I}$ by
\begin{equation}
h_{I}\equiv \tfrac{1}{3} \frac{\partial \mathcal{V} }{\partial h^{I}}\, , 
\end{equation}
\noindent
and are, therefore, homogenous of second degree in the $h^{I}$.

Clearly, as a function of the $h_{I}$, the function $\mathcal{V}$ is
homogeneous of degree $3/2$; this relation between the homogeneity degree of
the function $\mathcal{V}$ when expressed in different variables is a standard
implication of a Legendre transform, whence we define a new function
$\mathcal{W}(h_{\cdot})$ by
\begin{equation}
\mathcal{W}(h_{\cdot}) \; \equiv\; 3 h_{I}h^{I}(h_{\cdot}) -\mathcal{V}(h^{\cdot})
        \; =\; 2 \mathcal{V}(h_{\cdot})\, ,  
\end{equation}
\noindent
which is homogenous of degree $3/2$. The Legendre transform then immediately
implies that
\begin{equation}
h^{I}\; \equiv\; \tfrac{1}{3} \frac{\partial \mathcal{W} }{\partial h_{I}}\, .
\end{equation}

Next we introduce two new sets of variables $H_{I}$ and $\tilde{H}^{I}$,
related to the physical fields $(U,\phi^{x})$ by
\begin{equation}
  H_{I} \; \equiv\; e^{-U}h_{I}(\phi) \, ,\hspace{1cm}
  \tilde{H}^{I} \;\equiv \;  e^{-U/2}h^{I}(\phi) \, ,
\end{equation}
\noindent
and two new functions $\mathsf{V}$ and $\mathsf{W}$, which have the same form
in the new variables as $\mathcal{V}$ and $\mathcal{W}$ had in the old, {\it
  i.e.\/}
\begin{equation}
\mathsf{V}(\tilde{H})\,\equiv\,
C_{IJK}\tilde{H}^{I}\tilde{H}^{J}\tilde{H}^{K}\, ,
\hspace{1cm}
\mathsf{W}(H) \, \equiv\, 3\tilde{H}^{I}H_{I}\ -\ \mathsf{V}(\tilde{H})\ =\ 2\mathsf{V}\, ,
\end{equation}
but which are not constrained. These functions inherit the following properties
from $\mathcal{V}$ and $\mathcal{W}$:
\begin{eqnarray}
H_{I}  & \equiv & 
\tfrac{1}{3} \frac{\partial \mathsf{V} }{\partial  \tilde{H}^{I}}\, , 
\\
& & \nonumber \\  
\tilde{H}^{I} 
& \equiv & 
\tfrac{1}{3} \frac{\partial \mathsf{W} }{\partial H_{I}} \; \equiv\; \tfrac{1}{3}\ \partial^{I}\mathsf{W} \, . 
\end{eqnarray}
Using the homogeneity properties we find that
\begin{eqnarray}
e^{-\frac{3}{2}U} 
& = & 
\tfrac{1}{2} \mathsf{W}(H)\, ,
\\
& & \nonumber \\
h_{I} 
& = & 
(\mathsf{W}/2)^{-2/3}H_{I}\, , 
\\
& & \nonumber \\
h^{I} 
& = & 
(\mathsf{W}/2)^{-1/3}\tilde{H}^{I}\, . 
\end{eqnarray}

We can use these formulae to perform the change of variables in the FGK action
for static, spherically symmetric black holes of $N=2$, $d=5$ supergravity
\cite{Meessen:2011bd}, which in our conventions reads
\begin{equation}
\label{eq:effectived5h}
\mathcal{I}_{\text{FGK}}[U,\phi^{x}] 
= \int d\rho 
\left\{
(\dot{U})^{2}
+a^{IJ}\dot{h}_{I}\dot{h}_{J}
+e^{2U}a^{IJ}q_{I}q_{J}
+\mathcal{B}^{2}
\right\} .   
\end{equation}
It can be shown that 
\begin{equation}
a^{IJ} \; =\;  -\tfrac{2}{3} \left(\mathsf{W}/2 \right)^{4/3}\partial^{I}\partial^{J}
\log\, \mathsf{W}\, ,  
\end{equation}
\noindent
thus the above action, in terms of the $H_{I}$ variables, takes the form
\begin{equation}
\label{eq:effectived5H}
-\tfrac{3}{2}\mathcal{I}[H] 
= \int d\rho 
\left\{
\partial^{I}\partial^{J}
\log\, \mathsf{W} 
\left( \dot{H}_{I}\dot{H}_{J}+q_{I}q_{J}\right)
-\tfrac{3}{2}\mathcal{B}^{2}
\right\} .   
\end{equation}

The combination $\partial^{I}\partial^{J}\log\mathrm{W}$ appearing in the above
$\sigma$-model acts as a metric, so we are dealing with a mechanical problem
defined on a Hessian manifold.  As is well known, the $\rho$-independence of
the Lagrangian implies the conservation of the Hamiltonian $\mathcal{H}$. In
the FGK formalism, however, not all values of the energy are allowed and there
is a restriction called the Hamiltonian constraint. In the new variables this
constraint reads
\begin{equation}
\label{eq:hamiltoniand5}
\mathcal{H} \equiv   
\partial^{I}\partial^{J}
\log\, \mathsf{W} 
\left( \dot{H}_{I}\dot{H}_{J}-q_{I}q_{J}\right)
+\tfrac{3}{2}\mathcal{B}^{2}
=0\, .
\end{equation}
\noindent
The equations of motion derived from the effective action
(\ref{eq:effectived5H}) are\footnote{ The equations of motion can be obtained
  by taking the partial derivative of the Hamiltonian constraint with respect
  to $H_{K}$.  It goes without saying that having solved the Hamiltonian
  constraint does not imply having solved the equations of motion. }
\begin{equation}
\label{eq:eomsd5}
\partial^{K}\partial^{I}\partial^{J}
\log\, \mathsf{W} 
\left( \dot{H}_{I}\dot{H}_{J}-q_{I}q_{J}\right)
+2\partial^{K}\partial^{I}\log \mathsf{W}\, \ddot{H}_{I}
=0\, .
\end{equation}
Multiplying by $H_{K}$ and using the homogeneity properties of $\mathsf{W}$
and the Hamiltonian constraint we get
\begin{equation}
\label{eq:d5Ueq}
\partial^{I}\log \mathsf{W}\, \ddot{H}_{I} \; =\; \tfrac{3}{2}\mathcal{B}^{2}\, ,  
\end{equation}
\noindent
which is equivalent to the equation for the metric factor $U$ that one would
obtain from the action Eq.~(\ref{eq:effectived5h}) expressed in the new
variables. 

Eqs.~(\ref{eq:eomsd5}) are all the equations that need to be satisfied, but it
can be helpful to first solve the Hamiltonian constraint
(\ref{eq:hamiltoniand5}) or the equation of motion of $U$,
Eq.~(\ref{eq:d5Ueq}).

Observe that in the extremal case $\mathcal{B}=0$, the equations of motion can
be always satisfied by harmonic functions $\dot{H}_{I} = q_{I}$.


\section{H-FGK for \texorpdfstring{$N=2$}{N=2}, \texorpdfstring{$d=4$}{d=4} supergravity}
\label{sec-d4}

In the 4-dimensional case we also want to find a convenient change of
variables, from those defining a black-hole solution for given electric and
magnetic charges $(\mathcal{Q}^{M}) = (p^{\Lambda},q_{\Lambda})^{\mathrm{T}}$,
namely the metric function $U$ and the complex scalars $Z^{i}$, to the
variables $(H^{M}) = (H^{\Lambda},H_{\Lambda})^{\mathrm{T}}$ that have the same
transformation properties as the charges. There is an evident mismatch between
these two sets of variables, because $U$ is real. For consistency we will
introduce a complex variable $X$ of the form\footnote{ In this section we will
  be following the conventions of Ref.~\cite{Meessen:2006tu}, where the
  function $X$ appears as a scalar bilinear built out of the Killing spinors.  }
\begin{equation}
X = \tfrac{1}{\sqrt{2}}e^{U+i\alpha}\, ,
\end{equation}
\noindent
although the phase $\alpha$ does not occur in the original FGK formalism.  The
change of variables will then be well defined, and the absence of $\alpha$ will
lead to a constraint on the new set of variables: this constraint is related to
the absence of NUT charge, a possibility which in $d=4$ is allowed for by
spherical symmetry.

The theory is specified by the prepotential\footnote{We only use the
  prepotential here to determine quickly the homogeneity properties of the
  objects we are going to deal with. These properties are, however, valid for
  any $N=2$ theory in any symplectic frame, whether or not a prepotential
  exists.}  $\mathcal{F}$, a homogeneous function of second degree in the
complex coordinates $\mathcal{X}^{\Lambda}$. Consequently, defining
\begin{equation}
\mathcal{F}_{\Lambda} 
\equiv 
\frac{\partial\mathcal{F}}{\partial\mathcal{X}^{\Lambda}} \hspace{.5cm}\mbox{and}\hspace{.5cm}
\mathcal{F}_{\Lambda\Sigma} 
\equiv 
\frac{\partial^{2}\mathcal{F}}{\partial\mathcal{X}^{\Lambda}\partial\mathcal{X}^{\Sigma}}\,,
\hspace{.5cm}\mbox{we have:}\hspace{.5cm} 
\mathcal{F}_{\Lambda} = \mathcal{F}_{\Lambda\Sigma}\mathcal{X}^{\Sigma}\, .  
\end{equation}
\noindent
Since the matrix $\mathcal{F}_{\Lambda\Sigma}$ is homogenous of degree zero and
$X$ has the same K\"ahler weight as the covariantly holomorphic section
\begin{equation}
\left(\mathcal{V}^{M}\right) = 
\left(
  \begin{array}{c}
   \mathcal{L}^{\Lambda} \\
   \mathcal{M}_{\Lambda} \\
  \end{array}
\right)  
=
e^{\mathcal{K}/2}
\left(
  \begin{array}{c}
   \mathcal{X}^{\Lambda} \\
   \mathcal{F}_{\Lambda} \\
  \end{array}
\right),  
\end{equation}
\noindent
where $\mathcal{K}$ is the K\"ahler potential, we also find
\begin{equation}
\label{eq:MFL}
\frac{\mathcal{M}_{\Lambda}}{X} = \mathcal{F}_{\Lambda\Sigma}\, \frac{\mathcal{L}^{\Sigma}}{X}\, .
\end{equation}
\noindent
Defining the K\"ahler-neutral, real, symplectic vectors
$\mathcal{R}^{M}$ and $\mathcal{I}^{M}$ by
\begin{equation}
\mathcal{R}^{M}  = \Re\mathfrak{e}\, \mathcal{V}^{M}/X \, ,
\hspace{1cm}
\mathcal{I}^{M} = \Im\mathfrak{m}\,  \mathcal{V}^{M}/X\, , 
\end{equation}
\noindent
and using the symplectic metric 
\begin{equation}
\left(\Omega_{MN} \right) 
\equiv 
\left(
  \begin{array}{cc}
0 
& 
\mathbb{I}
\\
-\mathbb{I}
& 
0
\end{array}
\right)
\end{equation}
\noindent
as well as its inverse $\Omega^{MN}$ to lower and raise the symplectic indices
according to the convention
\begin{equation}
\mathcal{R}_{M} = \Omega_{MN}\mathcal{R}^{N}\, ,
\hspace{1cm}  
\mathcal{R}^{M} = \mathcal{R}_{N}\Omega^{NM}\, ,
\end{equation}
one can rewrite the complex relation (\ref{eq:MFL}) in the real form
\begin{equation}
\label{eq:RMI}
\mathcal{R}_{M} = -\mathcal{M}_{MN}(\mathcal{F})\mathcal{I}^{N}\, .
\end{equation}
\noindent
The symmetric symplectic matrix
\begin{equation}
\mathcal{M}(\mathcal{A}) 
\equiv 
\left(
  \begin{array}{cc}
\Im\mathfrak{m}\, \mathcal{A}_{\Lambda\Sigma} +
\Re\mathfrak{e}\, \mathcal{A}_{\Lambda\Omega}\, 
\Im\mathfrak{m}\, \mathcal{A}^{-1|\, \Omega\Gamma}\,
\Re\mathfrak{e}\, \mathcal{A}_{\Gamma\Sigma} 
& 
\hspace{.5cm}
-\Re\mathfrak{e}\, \mathcal{A}_{\Lambda\Omega}\,
\Im\mathfrak{m}\, \mathcal{A}^{-1 |\, \Omega\Sigma}
\\
\\
-
\Im\mathfrak{m}\, \mathcal{A}^{-1 |\, \Lambda\Omega}\,
\Re\mathfrak{e}\, \mathcal{A}_{\Omega\Sigma}
&
\Im\mathfrak{m}\, \mathcal{A}^{-1|\, \Lambda\Sigma}
\end{array}
\right),
\end{equation}
\noindent
can be associated with any symmetric complex matrix
$\mathcal{A}_{\Lambda\Sigma}$ with a non-degenerate imaginary part (such as
$\mathcal{F}_{\Lambda\Sigma}$ and the period matrix
$\mathcal{N}_{\Lambda\Sigma}$). The inverse of $\mathcal{M}_{MN}$, denoted by
$\mathcal{M}^{MN}$, is the result of raising the indices with the inverse
symplectic metric.

It is also immediate to prove the relation 
\begin{equation}
\label{eq:dRMdI}
d\mathcal{R}_{M} = -\mathcal{M}_{MN}(\mathcal{F})\, d\mathcal{I}^{N}\, .
\end{equation}
From this equality, its inverse and the symmetry properties of
$\mathcal{M}_{MN}$ we can derive the following relation between partial
derivatives (see {\it e.g.\/} \cite{Bellorin:2006xr}):
\begin{equation}
\label{eq:partials}
\frac{\partial \mathcal{I}^{M}}{\partial\mathcal{R}_{N}}  
=
\frac{\partial \mathcal{I}^{N}}{\partial\mathcal{R}_{M}}  
=
-\frac{\partial \mathcal{R}^{M}}{\partial\mathcal{I}_{N}}  
=
-\frac{\partial \mathcal{R}^{N}}{\partial\mathcal{I}_{M}}= -\mathcal{M}^{MN}(\mathcal{F}) \, .
\end{equation}

Similarly to what we did in five dimensions, we introduce two dual sets of
variables $H^{M}$ and $\tilde{H}_{M}$ and replace the original $n+1$ fields $X,
Z^{i}$ by the $2n+2$ real variables $H^{M}(\tau)$:
\begin{equation}
\mathcal{I}^{M}(X,Z,X^{*},Z^{*}) = H^{M}\, .  
\end{equation}
\noindent
The dual variables $\tilde{H}^{M}$ can be identified with $\mathcal{R}^{M}$,
which we can express as functions of the $H^{M}$ through
Eq.~(\ref{eq:RMI}). This gives $\mathcal{V}^{M}/X$ as a function of the
$H^{M}$. The physical fields can then be recovered by
\begin{equation}
Z^{i}\ =\ \frac{\mathcal{V}^{i}/X}{\mathcal{V}^{0}/X} \hspace{.7cm}\mbox{and}\hspace{.7cm}
e^{-2U}\ =\ \frac{1}{2|X|^{2}} \ =\ \mathcal{R}_{M}\mathcal{I}^{M}\, .  
\end{equation}
\noindent
The phase of $X$, $\alpha$, can be found by solving the differential equation
({\it cf.\/}~Eqs.~(3.8), (3.28) in Ref.~\cite{Galli:2010mg})
\begin{equation}
\dot{\alpha} =2|X|^{2} \dot{H}^{M}H_{M} -\mathcal{Q}_{\star}\, ,
\hspace{.5cm}\mbox{where}\hspace{.5cm}
\mathcal{Q}_{\star} = \tfrac{1}{2i}\dot{Z}^{i}\partial_{i}\mathcal{K} +\mathrm{c.c.} 
\end{equation}
\noindent
is the pullback of the K\"ahler connection 1-form
\begin{equation}
\mathcal{Q}_{\star} = \tfrac{1}{2i}\dot{Z}^{i}\partial_{i}\mathcal{K} +\mathrm{c.c.}  
\end{equation}

Having detailed the change of variables, we want to rewrite the FGK action for
static, spherically symmetric solutions of $N=2$, $d=4$ supergravity
\cite{Ferrara:1997tw}, {\it i.e.\/}
\begin{equation}
\label{eq:effectiveaction}
I_{\text{FGK}}[U,Z^{i}] = \int d\tau \left\{ 
(\dot{U})^{2}  
+\mathcal{G}_{ij^{*}}\dot{Z}^{i}  \dot{Z}^{*\, j^{*}}  
-\tfrac{1}{2}e^{2U}
\mathcal{M}_{MN}(\mathcal{N})\mathcal{Q}^{M}\mathcal{Q}^{N} 
+r_{0}^{2}
\right\} ,  
\end{equation}
\noindent
in terms of the variables $H^{M}$. 
As in the 5-dimensional case, we start by defining the function
$\mathsf{W}(H)$
\begin{equation}
\mathsf{W}(H) \equiv \tilde{H}_{M}(H)H^{M} = e^{-2U} = \frac{1}{2|X|^{2}}\, ,
\end{equation}
\noindent
which is homogenous of second degree in the $H^{M}$. Using the properties
(\ref{eq:partials}) one can show that
\begin{eqnarray}
\partial_{M}\mathsf{W}
& \equiv & 
\frac{\partial\mathsf{W}}{\partial H^{M}}
= 
2\tilde{H}_{M}\, , 
\\
& & \nonumber \\
\partial^{M}\mathsf{W}
& \equiv & 
\frac{\partial\mathsf{W}}{\partial \tilde{H}_{M}}
= 
2H^{M}\, , 
\\
& & \nonumber \\
\partial_{M}\partial_{N}\mathsf{W} 
& = & 
-2\mathcal{M}_{MN}(\mathcal{F})\, ,
\\
& & \nonumber \\
\mathsf{W}\,\partial_{M}\partial_{N}\log \mathsf{W} 
& = & 
2\mathcal{M}_{MN}(\mathcal{N})+4\mathsf{W}^{-1}H_{M}H_{N}\, ,
\end{eqnarray}
\noindent
where the last property is based on the following relation\footnote{ This
  relation can be derived from the identities in Ref.~\cite{Ceresole:1995ca}.
}
\begin{equation}
-\mathcal{M}_{MN}(\mathcal{N})  
=
\mathcal{M}_{MN}(\mathcal{F})  
+4\mathcal{V}_{(M}\mathcal{V}^{*}_{N)}\, .
\end{equation}

Using the special geometry identity $\mathcal{G}_{ij^{*}} =
-i\mathcal{D}_{i}\mathcal{V}_{M}\mathcal{D}_{j^{*}}\mathcal{V}^{*\, M}$, we
can rewrite the effective action in the form
\begin{equation}
\label{eq:effectiveaction2}
-I_{\rm eff}[H] 
= 
\int d\tau 
\left\{ 
\tfrac{1}{2}\partial_{M}\partial_{N}\log\mathsf{W} 
\left(\dot{H}^{M}\dot{H}^{N}+\tfrac{1}{2}\mathcal{Q}^{M}\mathcal{Q}^{N} \right) 
-\Lambda
-r_{0}^{2}
\right\} ,  
\end{equation}
\noindent
where we have defined
\begin{equation}
\Lambda \equiv \left(\frac{\dot{H}^{M}H_{M}}{ \mathsf{W}}\right)^{2} 
+\left(\frac{\mathcal{Q}^{M}H_{M}}{ \mathsf{W}}\right)^{2}\, .
\end{equation}
\noindent
The $\tau$-independence of the Lagrangian implies the conservation of the
Hamiltonian $\mathcal{H}$
\begin{equation}
\mathcal{H}\equiv 
-\tfrac{1}{2}\partial_{M}\partial_{N}\log\mathsf{W} 
\left(\dot{H}^{M}\dot{H}^{N}-\tfrac{1}{2}\mathcal{Q}^{M}\mathcal{Q}^{N} \right) 
+\left(\frac{\dot{H}^{M}H_{M}}{ \mathsf{W}}\right)^{2} 
-\left(\frac{\mathcal{Q}^{M}H_{M}}{ \mathsf{W}}\right)^{2}
-r_{0}^{2}=0\, .  
\end{equation}

The equations of motion can be written in the form
\begin{equation}
\tfrac{1}{2}\partial_{P}\partial_{M}\partial_{N}\log \mathsf{W} 
\left(\dot{H}^{M}\dot{H}^{N} -\tfrac{1}{2}\mathcal{Q}^{M}\mathcal{Q}^{N}
\right)  
+\partial_{P}\partial_{M}\log \mathsf{W}\, \ddot{H}^{M}
-\frac{d}{d\tau}\left(\frac{\partial \Lambda}{\partial \dot{H}^{P}}\right)
+\frac{\partial \Lambda}{\partial H^{P}}=0\, .
\end{equation}
\noindent
Contracting them with $H^{P}$ and using the homogeneity properties of the
different terms as well as the Hamiltonian constraint above, we find the
equation ({\it cf.\/}~Eq.~(3.31) of Ref.~\cite{Galli:2010mg} for the stationary
extremal case)
\begin{equation}
\label{eq:Urewriten}
\tfrac{1}{2}\partial_{M}\log \mathsf{W} \left(\ddot{H}^{M}
  -r_{0}^{2}H^{M}\right) +\left(\frac{\dot{H}^{M}H_{M}}{ \mathsf{W}}\right)^{2}  = 0\, ,
\end{equation}
\noindent
which corresponds to the equation of motion of the variable $U$ in the standard
formulation.

Note that in the extremal case ($r_{0}=0$) and in the absence of the NUT charge
($\dot{H}^{M}H_{M}=0$) the equations of motion are solved by harmonic functions
$\dot{H}^{M} = \mathcal{Q}^{M}$ \cite{Bellorin:2006xr}.


\section*{Acknowledgments}

This work has been supported in part by the Spanish Ministry of Science and
Education grant FPA2009-07692, a Ram\'on y Cajal fellowship RYC-2009-05014,
the Princip\'au d'Asturies grant IB09-069, the Comunidad de Madrid grant
HEPHACOS S2009ESP-1473, and the Spanish Con\-so\-lider-Ingenio 2010 program CPAN
CSD2007-00042. The work of C.S.S.~has been supported by a JAE-predoc grant JAEPre
2010 00613. T.O.~wishes to thank M.M.~Fern\'andez for her ferrous support.


\end{document}